\pdfoutput=1
\documentclass[10pt, conference, compsocconf]{IEEEtran}
\usepackage{slashbox}
\usepackage{xspace}
\usepackage{graphicx}

\ifCLASSINFOpdf
\else
\fi

\begin{document}
%
\title{RFP: A Remote Fetching Paradigm for RDMA-Accelerated Systems}

\author{\IEEEauthorblockN{Maomeng Su\IEEEauthorrefmark{1},
Mingxing Zhang\IEEEauthorrefmark{1},
Kang Chen\IEEEauthorrefmark{2},
Yongwei Wu\IEEEauthorrefmark{2}, and
Guoliang Li\IEEEauthorrefmark{2}}
\IEEEauthorblockA{Department of Computer Science and Technology, Tsinghua University, China}
\IEEEauthorblockA{\IEEEauthorrefmark{1}\{smm11,zhangmx12\}@mails.tsinghua.edu.cn~~\IEEEauthorrefmark{2}\{chenkang,wuyw,liguoliang\}@tsinghua.edu.cn}
}


\maketitle

\newcommand{\reminder}[1]{ {\mbox{$<==$}} [[[ { \bf #1 } ]]] {\mbox{$==>$}}}

\newcommand{\Mem}{\texttt{Memcached}\xspace}
\newcommand{\Redis}{\texttt{Redis}\xspace}

\newcommand{\sr}{ServerReply\xspace}
\newcommand{\Jakiro}{Jakiro\xspace}
\newcommand{\osu}{RDMA-Memcached\xspace}
\newcommand{\Pilaf}{Pilaf\xspace}

\newtheorem{theorem}{Theorem}
\newtheorem{example}{Example}
\newtheorem{definition}{Definition}
\newtheorem{proposition}{Proposition}
\newtheorem{lemma}{Lemma}


\begin{abstract}
Remote Direct Memory Access (RDMA) is an efficient way to improve the performance of traditional client-server systems. Currently, there are two main design paradigms for RDMA-accelerated systems. The first allows the clients to directly operate the server's memory and totally bypasses the CPUs at server side. The second follows the traditional server-reply paradigm, which asks the server to write results back to the clients. However, the first method has to expose server's memory and needs tremendous re-design of upper-layer software, which is complex, unsafe, error-prone, and inefficient. The second cannot achieve high input/output operations per second (IOPS), because it employs out-bound RDMA-write at server side which is not efficient.

We find that the performance of out-bound RDMA-write and in-bound RDMA-read is asymmetric and the latter is 5 times faster than the former. Based on this observation, we propose a novel design paradigm named Remote Fetching Paradigm (RFP). In RFP, the server is still responsible for processing requests from the clients. However, counter-intuitively, instead of sending results back to the clients through out-bound RDMA-write, the server only writes the results in local memory buffers, and the clients use in-bound RDMA-read to remotely fetch these results. Since in-bound RDMA-read achieves much higher IOPS than out-bound RDMA-write, our model is able to bring higher performance than the traditional models.

In order to prove the effectiveness of RFP, we design and implement an RDMA-accelerated in-memory key-value store following the RFP model. To further improve the IOPS, we propose an optimization mechanism that combines status checking and result fetching. Experiment results show that RFP can improve the IOPS by 160\%$\sim$310\% against state-of-the-art models for in-memory key-value stores.
\end{abstract}



%
\IEEEpeerreviewmaketitle


\section{Introduction}
Recently, high-performance interconnect solutions, such as InfiniBand, have been prevalently deployed in commodity datacenters for their large bandwidth (e.g., 40, 56, or even 100 Gbps) \cite{jakiro_farm}. Nevertheless, due to the dominance of small packets in real client-server applications (e.g., key-value stores) \cite{jakiro_facebook,jakiro_smalldata}, the input/output operations per second (IOPS), rather than the maximum bandwidth, becomes the real bottleneck in commodity datacenters. To alleviate this problem, the Remote Direct Memory Access (RDMA) technology is revisited in an InfiniBand environment by many previous studies \cite{jakiro_chint,jakiro_farm,jakiro_pilaf,jakiro_mpich2rdma,jakiro_wimpy,jakiro_memcachedrdma}. Because of the simplification of protocol stack and the CPU-bypassing ability, RDMA can achieve much higher IOPS than the traditional TCP/IP. However, to the best of our knowledge, the potential benefits of RDMA over InfiniBand Network Interface Card (INIC) has not been fully explored and there leaves an opportunity to further improve the IOPS.

There are currently two main design paradigms for using RDMA in client-server systems. The first is the server-bypass model. It allows the clients to directly read/write server's memory through RDMA and bypasses CPUs at the server side \cite{jakiro_farm,jakiro_chint,jakiro_pilaf}. However, since the server is totally unaware of the request operations, this totally-bypass design paradigm is always specialized for a certain purpose and cannot be generalized to arbitrary client-server applications (e.g., RPC services). Even worse, as the server is not involved in data processing, it should explicitly expose all critical memory \footnote{Critical memory denotes the memory region that stores core data such as key-value items in a system, which is usually in a large amount (GB or TB).} to clients \cite{jakiro_farm}. Such memory is used to store data and could be dynamically allocated and reclaimed by the server. Thus, clients have to rely on themselves to prevent from accessing data that have been invalidated by the server \cite{jakiro_chint}. This model also has the data consistency issues as both the server CPU and clients can access the memory. In summary, the totally-bypass paradigm needs tremendous re-design of upper-layer software, and is complex, unsafe, error-prone, and inefficient.

The second paradigm employs the traditional server-reply model, in which the server is responsible for processing requests from the clients and sends results back to them through RDMA-write \cite{jakiro_herd,jakiro_memcachedrdma,jakiro_mpich2rdma,jakiro_pvfs,jakiro_hbase}. Just like the usage of TCP/IP, under this circumstance, the server does not need to expose its critical memory to the clients, which avoids data races between the server and the clients. Moreover, as the server is involved in processing procedure, it can execute arbitrary management policy that is transparent to the clients. Thus the server-reply design paradigm is general enough to support nearly all applications. However, it relies on out-bound RDMA-write to send results back to clients which is not efficient, and thus has low performance in IOPS.





We find that the hardware design principle determines that the INIC performs much differently in IOPS between \emph{in-bound} RDMA and \emph{out-bound} RDMA, where the IOPS of out-bound RDMA indicates the number of RDMA operations an INIC can issue to other INICs per second, and the IOPS of in-bound RDMA stands for the number of RDMA operations an INIC can serve per second. As more states and operations are maintained for an INIC to issue RDMA operations \cite{jakiro_herd}, the IOPS of out-bound RDMA is much lower than the IOPS of in-bound RDMA. According to our test in an InfiniBand-based cluster, the peak in-bound RDMA IOPS is 11.26 MOPS (million operations per second) of an INIC (MT27500, 40 Gbps) while the the peak out-bound RDMA IOPS is only 2.11 MOPS. Thus in-bound RDMA is 5 times faster than out-bound RDMA. Consequently, the traditional server-reply paradigm, whose maximum IOPS is bounded by server's out-bound RDMA rather than in-bound RDMA, is not the best choice for transferring small requests.

To address this problem, we propose a novel design paradigm, \textbf{\emph{Remote Fetching Paradigm}} (RFP). Similar to the traditional server-reply design paradigm, RFP asks the clients to send requests to the server by using RDMA-write and the server to process these requests. However, counter-intuitively in RFP, instead of sending the results or notifications back to the clients through (out-bound) RDMA-write, the server only reserves the results in local memory buffers. It is the clients that use in-bound RDMA-read to remotely fetch results from these memory buffers. The unique feature of RFP by offloading the responsibility of transferring results from the server to the clients is important to achieve higher IOPS, because it can make the most of the in-bound RDMA performance of the server's INIC. As the server does not participate in the networking operations, the limitation on out-bound RDMA is avoided.

In order to evaluate the effectiveness of our RFP model, we have built an RDMA-based in-memory key-value store named Jakiro. As shown by many previous works, the in-memory key-value store, which has been widely used as a backend system in many data centers, is a good representative for evaluating the usage of RDMA \cite{jakiro_chint,jakiro_pilaf,jakiro_herd,jakiro_memcachedrdma}.

In summary, we make the following contributions.
\begin{enumerate}
\item We propose a novel design paradigm named Remote Fetching Paradigm (RFP), which not only is a general and safe solution to support all client-server applications, but also makes the most of in-bound RDMA over INIC to achieve high IOPS on small data. RFP outperforms the totally-bypass design paradigm by taking advantage of making server CPUs involved in critical process logic, and outperforms the server-reply design paradigm by allowing clients to fetch the results actively and remotely instead of asking the server to reply them. We believe that this new paradigm can help practitioners rethink the design of traditional systems or applications over RDMA.
\item We have built Jakiro, an RDMA-based in-memory key-value store, based on RFP. We propose an optimization mechanism by combining checking result status and fetching results. The result is encapsulated into the data region which will be transferred to the client during the status-checking action. Hence, one RDMA-read operation is enough for both status and results, which can further improve the IOPS.
\item Experiment results show that compared to in-memory key-value stores that adopt totally-bypass and server-reply design paradigms, Jakiro improves the IOPS by 160\%$\sim$310\% under different (uniform, skew, and dynamic) workloads. Jakiro also reduces the latency by more than 50\% on small data.
\end{enumerate}

The rest of this paper is organized as follows. In Section~\ref{background}, we give the background of InfiniBand and RDMA, as well as in-memory key-value stores. We propose Remote Fetching Paradigm (RFP) in Section~\ref{remote_fetch}. Section~\ref{design} briefly describes the design of Jakiro following RFP. In Section~\ref{evaluation}, we provide the experimental results of Jakiro. Section~\ref{related_work} reviews related work and Section~\ref{conclusion} concludes our work.



\section{Background}\label{background}



\subsection{InfiniBand and RDMA}
InfiniBand has been used in high performance computing for decades \cite{jakiro_mpi,jakiro_mpich2rdma,jakiro_pvfs,jakiro_mallanox}. The InfiniBand hardware can offer bandwidth at 20, 40, 56, and even 100 Gbps. Thanks to the price dropping of network interface cards and switches, InfiniBand is being deployed in more commodity data centers. There are three main communication protocols supported in InfiniBand: IPoIB (IP over InfiniBand), Send/Recv verbs, and Remote Direct Memory Access (RDMA) verbs. As reported in previous works \cite{jakiro_farm,jakiro_chint,jakiro_pilaf,jakiro_mpi}, both IPoIB and Send/Recv verbs (in reliable mode) cannot outperform RDMA verbs on small data ($\sim$32 bytes). Thus in this paper, we only focus on the performance of RDMA.

RDMA enables the CPU in one machine to directly read/write the memory in another remote machine, without involving the CPU and the OS kernel of the remote machine. Hence, RDMA is also called as one-sided operation: only the CPU in the client machine is aware of the access. Meanwhile, RDMA avoids the overhead of copying data between the user space and the kernel space. RDMA-read and RDMA-write are two main RDMA verbs used.

\subsection{In-Memory Key-Value Store}

In-memory key-value store has been widely deployed in current data centers, e.g., Facebook \cite{jakiro_facebook}. It is used either as a storage service (e.g., RAMCloud \cite{jakiro_ramcloud}) to store the data in memory, or as a cache middleware (e.g., Memcached \cite{jakiro_memcached}) to accelerate the operation performance of underlying disk-based storage. Most existing RDMA-based systems, e.g., Pilaf \cite{jakiro_pilaf}, C-Hint \cite{jakiro_chint}, and FaRM \cite{jakiro_farm}, have used in-memory key-value store as a test case to evaluate the RDMA-accelerated techniques.


In real cases, small items dominate most key-value stores \cite{jakiro_smalldata}. According to the analysis of the work in \cite{jakiro_facebook}, the value size of more than half of key-value items in Facebook's datacenter is around 20 bytes. This is common and reasonable for most applications. For example, for an examination application, the value to record the examination point would be a 4-byte float variable. The MD5 code \cite{jakiro_md5}, which is widely used by many applications for checksum, uses 16 bytes to store the coding result. For those small key-value items, IOPS should be more cared about and optimized. In this paper, we also build an in-memory key-value store to prove the effectiveness of our model, which is proposed for achieving high IOPS.


\section{Remote Fetching Paradigm for RDMA}\label{remote_fetch}

In this section, we first present the performance difference between out-bound RDMA and in-bound RDMA in terms of IOPS. Then we define the Remote Fetching Paradigm to make the most of the inherent features of RDMA.

\subsection{Out-bound and In-bound RDMA}

We first give the definitions of out-bound and in-bound RDMA operation.

\textbf{Out-bound RDMA-read/write.} An RDMA-read/write issued from an INIC (to other INICs) is called an out-bound RDMA-read/write w.r.t. the INIC.

\textbf{In-bound RDMA-read/write.} An RDMA-read/write served by an INIC (issued from other INICs) is called an in-bound RDMA-read/write w.r.t. the INIC.

For a server program, it has two choices for communication with clients. One is through its out-bound RDMA and the other is through its in-bound RDMA. For out-bound RDMA, the server initiates RDMA operations and writes to (reads from) multiple clients. For in-bound RDMA, the clients initiate the operations and write to (read from) the single server. However, the performance of in-bound RDMA and out-bound RDMA in the server are asymmetric. The performance is evaluated by in-bound and out-bound RDMA IOPS respectively, which is defined as below.

\textbf{Out-bound RDMA IOPS.} The out-bound RDMA-read/write IOPS of an INIC is the number of out-bound RDMA-read/write operations it issues per second.

\textbf{In-bound RDMA IOPS.} The in-bound RDMA-read/write IOPS of an INIC is the number of in-bound RDMA-read/write operations it serves per second.

Next we evaluate the asymmetric IOPS of in-bound and out-bound RDMA in real environment.

\subsection{Asymmetric IOPS of Out-bound and In-bound RDMA}


We use a cluster of eight machines to test the difference between out-bound RDMA-write and in-bound RDMA-read. Each machine is equipped with a Mallanox ConnectX-3 INIC (MT27500, 40 Gbps) and dual 8-core CPUs (Intel Xeon E5-2640 v2, 2.0 GHz). The detailed information of these machines will be presented in Section 5.

In the evaluation, we choose one machine to be the server machine and other seven machines to be the client machines. This is a typical client-server architecture. We do not run any service on these machines, but purely test the IOPS of RDMA-write and RDMA-read on small data. The out-bound RDMA-write IOPS is tested by making the server machine continuously issuing RDMA-write operations to other seven client machines. We launch a number of threads in the server machine, each connected by threads in the client machines. Each server thread randomly chooses a client thread and issues an RDMA-write operation to it, and repeats such action after this operation is completed. The in-bound RDMA-read IOPS is tested by making the seven client machines continuously issue RDMA-read operations to the server machine. We launch a number of client threads in the seven client machines, each connecting to the server machine. As the server machine is unaware of the RDMA-read operations, the number of server threads have no impact on the IOPS of in-bound RDMA-read. Each client thread has its own memory buffer in the server machine, and it only repeats issuing RDMA-read operations to the memory buffer belonging to it.

It has to be mentioned that the \emph{inlined} mechanism is applied to optimize RDMA-write in the test. The inlined mechanism directly inlines data into the request in the INIC, without incurring DMA (Direct Memory Access) operations. It brings lower latency and higher IOPS on small data than the unlined mechanism \cite{jakiro_herd,jakiro_mpi,jakiro_pvfs}, which needs a DMA to fetch the data from the memory. However, the inlined mechanism cannot be used for RDMA-read.

\begin{figure}[!t]
\begin{minipage}{0.45\linewidth}
  \centerline{\includegraphics[width=1.9in]{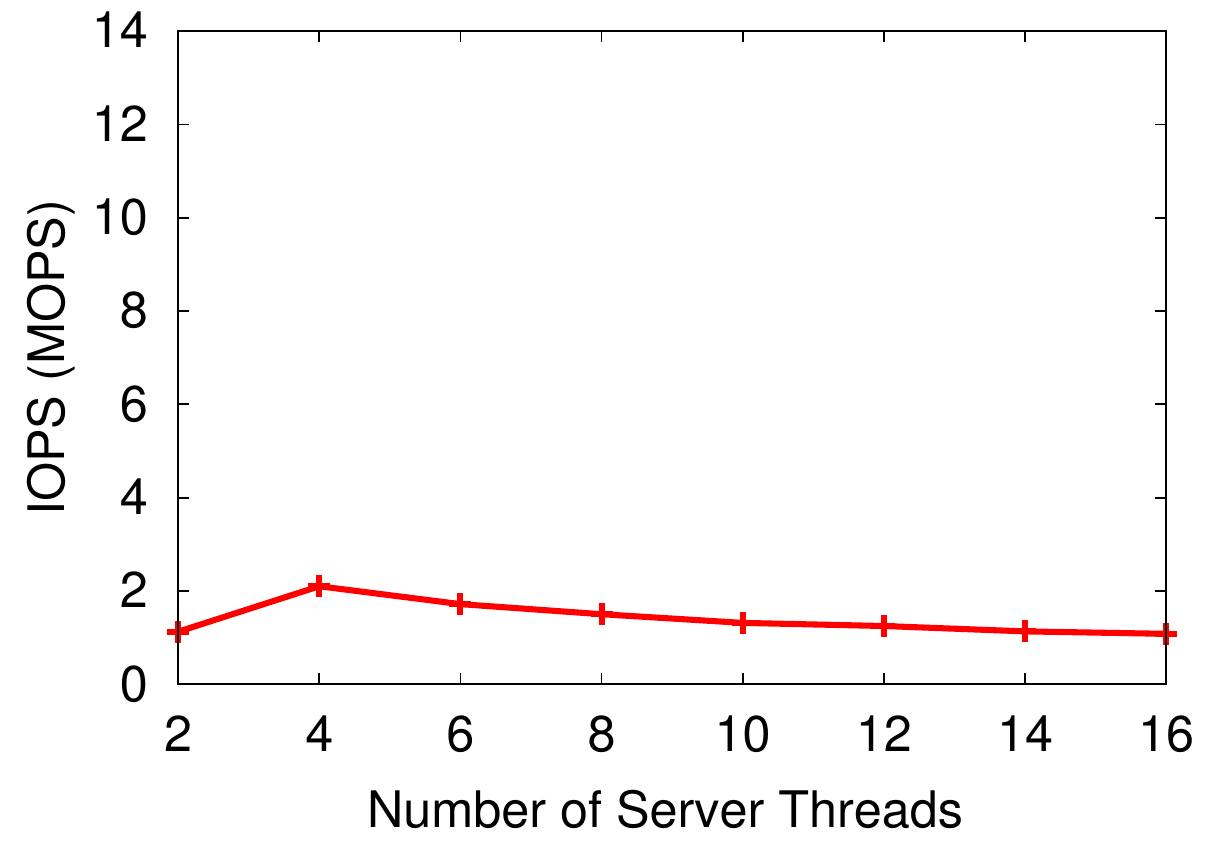}}
  \centerline{(a) Out-bound RDMA-write}
\end{minipage}
\hfill
\begin{minipage}{0.45\linewidth}
  \centerline{\includegraphics[width=1.9in]{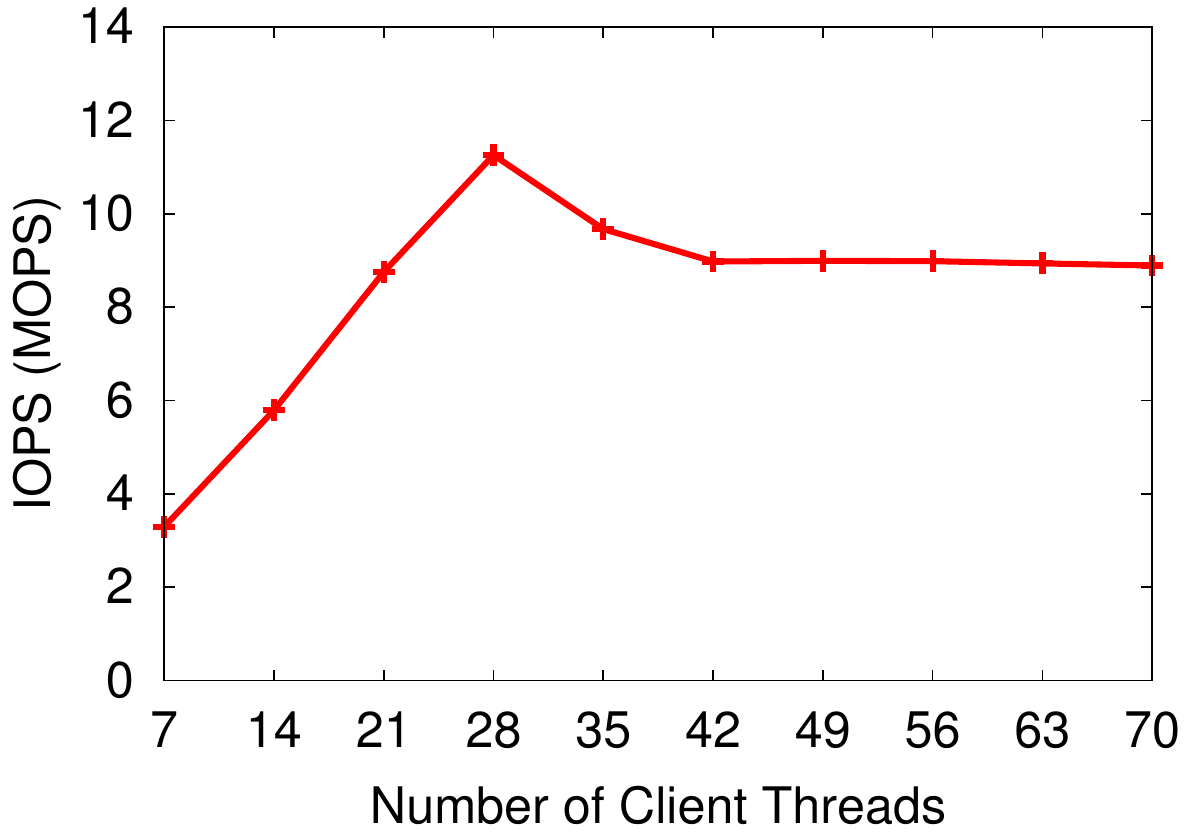}}
  \centerline{(b) In-bound RDMA-read}
\end{minipage}
\caption{The IOPS for out-bound RDMA and in-bound RDMA on 32-byte data items. The client threads are uniformly distributed among seven client machines.}
\label{j_iops_dif}
\vspace{-1em}
\end{figure}


Figure~\ref{j_iops_dif} illustrates the IOPS difference between out-bound RDMA-write and in-bound RDMA-read on 32-byte data items. A key observation can be made from Figure~\ref{j_iops_dif}:

\vspace{.5em}
\noindent \fbox{\shortstack[l]{\textbf{Observation.} To an INIC, the peak IOPS of in-bound \\ RDMA-read (11.26 MOPS) is about 5X as high as that \\ of out-bound RDMA-write (2.11 MOPS).}}
\vspace{.5em}



The IOPS of out-bound RDMA-write (in the order of two millions) is much lower than that of in-bound RDMA-read (in the order of ten millions). The reason for this phenomenon is that when the local INIC issues an RDMA request to another remote INIC, the local INIC at least has to prepare the request and its corresponding states in the hardware, send the request, wait for an ACK/NACK notification from the remote INIC, and generate an event when the request is completed \cite{jakiro_herd}. These operations require the local INIC to maintain a number of states in the hardware, which significantly limits the capacity of the local INIC to perform more RDMA operations. Figure~\ref{j_iops_dif}(a) gives the IOPS of out-bound RDMA-write under different numbers of server threads. Four server threads achieve 2.11 MOPS, which is higher than other numbers of server threads. Figure~\ref{j_iops_dif}(a) also shows that out-bound RDMA-write imposes poor scalability. This is because when the number of server threads increases, the threads will compete for the shared resources in the INIC, which constrains the scalability and IOPS.

The results in Figure~\ref{j_iops_dif}(a) are tested by making each server thread connected by seven client threads, which are uniformly distributed among the client machines. We also test the out-bound RDMA-write IOPS with each server thread connected by more client threads per machine. As the server has to use more hardware resources to maintain more connections, IOPS in these cases would be no higher than that of maintaining less connections. Thus, we do not plot these results in Figure~\ref{j_iops_dif}(a).

\begin{table}[!t]
\renewcommand{\arraystretch}{1.3}
\caption{The IOPS (measured in MOPS) on 32-byte data items for out-bound RDMA-write under different client machine numbers. Each machine runs one client thread, connecting to all the server threads. }
\vspace{-0.5em}
\label{table_machine_number}
\centering
\begin{tabular}{|c|c|c|c|c|}
\hline
\backslashbox{\#ServerThreads}{\#Machine} & 1 & 3 & 5 & 7\\
\hline
1 & 0.62 & 0.61 & 0.61 & 0.61\\
\hline
2 & 1.12 & 1.13 & 1.12 & 1.12\\
\hline
4 & 2.17 & 2.13 & 2.11 & 2.11\\
\hline
8 & 1.46 & 1.46 & 1.51 & 1.50\\
\hline
\end{tabular}
\vspace{-1em}
\end{table}

\begin{table}[htup]
\caption{IOPS comparison (MOPS) of CPU-read/write, in-bound RDMA-read/write, and out-bound RDMA-read/write on 32-byte data.}
\vspace{-0.5em}
\label{table_cpuinout}
\centering
\begin{tabular}{|l|c|c|c|}
\hline
CPU-read/write (using a single core) & around 41.9 \\
\hline
In-bound RDMA-read/write & around 11.3\\
\hline
Out-bound RDMA-read/write & around 2.1 \\
\hline
\end{tabular}
\end{table}

Moreover, the IOPS bottleneck of out-bound RDMA-write is not impacted by the number of client machines. For example, keeping four threads in the server machine, we vary the client machine number from 1 to 7. We find that out-bound RDMA-write is still bounded at the order of two millions, as shown in Table~\ref{table_machine_number}.

However, for in-bound RDMA-read, the INIC in the server machine has little processing burden but only  serve the incoming requests. In this case, the INIC does not have to maintain much state or do extra operations for in-bound requests. The capacity of the INIC can be fully utilized and thus brings higher IOPS. We can see from Figure~\ref{j_iops_dif}(b) that the INIC of the server machine can achieve peak IOPS at 11.26 MOPS for in-bound RDMA-read on 32-byte data items. It is about 5x as high as that of out-bound RDMA-write. Such in-bound RDMA-read IOPS is similar to that reported in FaRM \cite{jakiro_farm} and C-Hint \cite{jakiro_chint}.

The in-bound RDMA-read IOPS of the server machine decreases when more client threads are launched in each client machine. This is because RDMA-read requests issued by the client machine's INIC are limited by the increasing number of client threads, the same reason as out-bound RDMA-write. Note that when the data size grows larger than 1.5 KB, in-bound RDMA and out-bound RDMA perform the same in IOPS, as the bandwidth becomes the bottleneck. When data size is less than 1.5 KB, in-bound RDMA significantly outperforms out-bound RDMA in IOPS, which is the bottleneck at this time.

Table~\ref{table_cpuinout} further compares IOPS on 32-byte data among CPU-read/write, in-bound RDMA-read/write, and out-bound RDMA-read/write. CPU-read/write outperforms in-bound RDMA-read/write on small data, which is better than out-bound RDMA-read/write, as listed in Table~\ref{table_cpuinout} \footnote{The INIC with 20 Gbps has the same results, and we do not discuss them further.}.

According to the discussion in this section, the traditional wisdom that makes the server send replies back to clients will suffer from the bottleneck of INIC's out-bound RDMA-write. However, there leaves a great opportunity to make the most of in-bound RDMA-read performance to optimize the design of client-server systems.


\subsection{New Design Paradigm of Using In-bound RDMA}

\begin{figure}[!t]
\centering
\includegraphics[width=2.8in,height=2.4in]{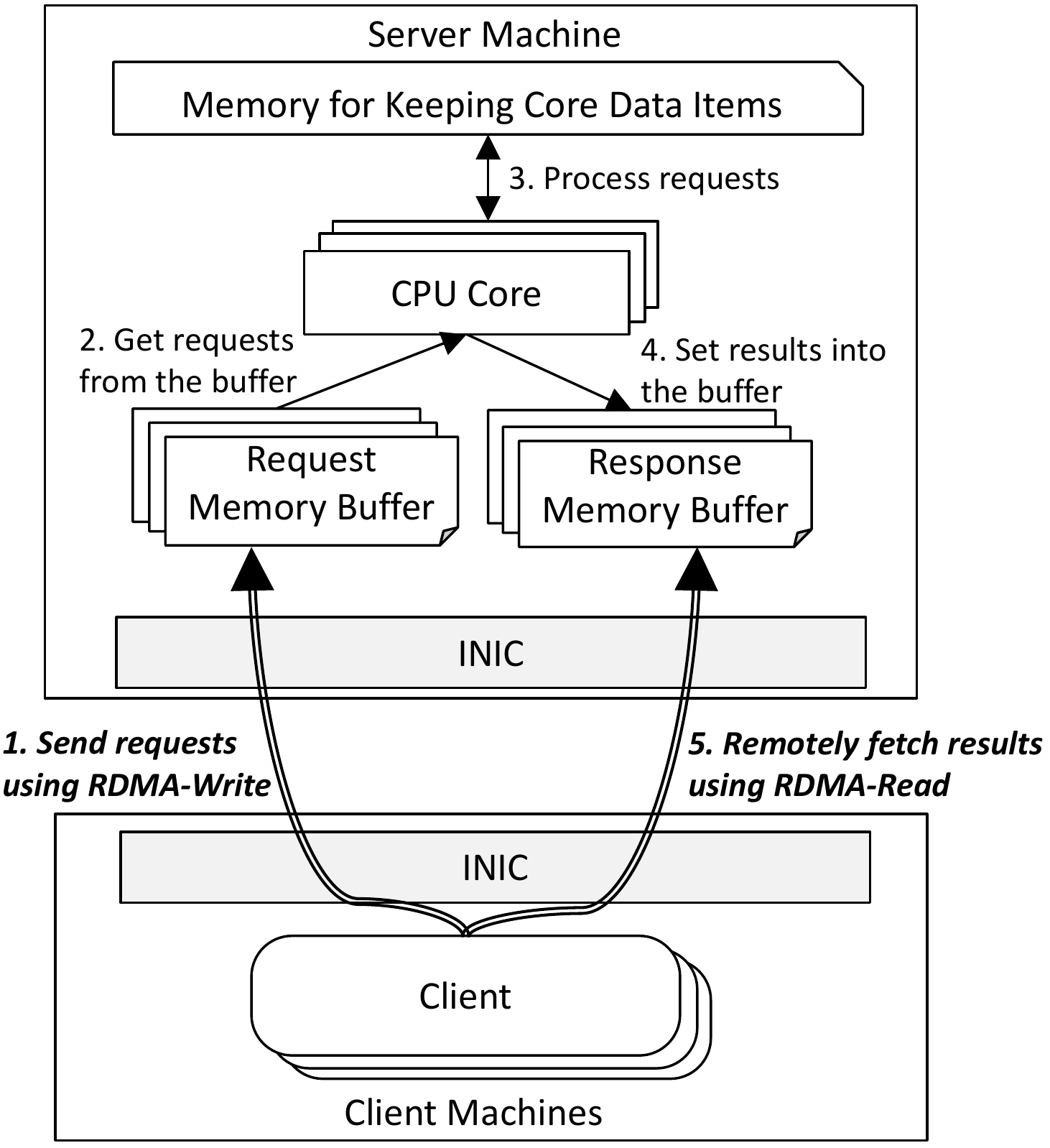}
\caption{Remote Fetching Paradigm (RFP) of applying RDMA in system or application design. The unique features of RFP are Step 1 and Step 5, which use in-bound RDMA of server's INIC.}
\vspace{-1em}
\label{j_arch}
\end{figure}

Based on the observation of the IOPS difference between out-bound RDMA-write and in-bound RDMA-read, we present Remote Fetching Paradigm (RFP), a novel paradigm for client-server system aiming to achieve excellent IOPS on small data. Partly like the server-reply design paradigm, RFP still asks the clients to use RDMA-write to send their requests and the server to process these requests. However, as illustrated in Figure~\ref{j_arch}, in RFP, the server does not send any result or notification back to clients through RDMA-write after it successfully processes the requests, but writes these results into local memory buffers. The clients use RDMA-read to remotely check and fetch the results from these buffers. Table~\ref{table_comapre3} lists the comparisons among totally-bypass paradigm, server-reply paradigm, and RFP.

RFP has several salient features. Firstly, as RFP also relies on the server to process the requests, it maintains the general feature of the traditional server-reply paradigm to support most types of systems or applications. Secondly, as displayed in Figure~\ref{j_arch}, the server does not have to expose all memory regions to the clients. It only exposes a small number of request/response memory buffers to keep requests and results. Thus, the critical memory that keeps core data items can be protected and operated by the server locally. This avoids the races between clients and the server existing in totally-bypass design paradigm \cite{jakiro_pilaf,jakiro_chint,jakiro_farm}. Clients also do not need to reason about data consistency. Moreover, each client has its own request/response buffers at the server side. They are not authorized to access buffers not belonging to them. Finally and most importantly, RFP does not need to waste server CPU cycles on operations of sending results back through the network, which is different from the traditional wisdom. Instead, the server CPUs only write the results into local response buffers. This could eliminate the bottleneck of out-bound RDMA-write on small data at the server side. RFP allows clients to actively and remotely fetch the results using RDMA-read, and thus is able to fully utilize high in-bound RDMA-read IOPS of the server's INIC.

\begin{table}[!t]
\renewcommand{\arraystretch}{1.3}
\caption{Comparison of totally-bypass design paradigm, server-reply design paradigm, and RFP.}
\vspace{-0.5em}
\label{table_comapre3}
\centering
\begin{tabular}{|p{3cm}|c|c|c|}
\hline
 & Totally-Bypass & Server-Reply & RFP\\
\hline
Server exposing critical memory & $\surd$ & \large{$\times$} & \large{$\times$}\\
\hline
Server processing requests & \large{$\times$} & $\surd$ & $\surd$\\
\hline
Client processing requests & $\surd$ & \large{$\times$} & \large{$\times$}\\
\hline
Server sending back results & \large{$\times$} & $\surd$ & \large{$\times$}\\
\hline
Clients fetching results remotely & \large{$\times$} & \large{$\times$} & $\surd$\\
\hline
\end{tabular}
\vspace{-1em}
\end{table}

Besides writing result payloads into local response buffers, the server has to attach additional information to indicate whether the results are ready or not. The clients use RDMA-read to remotely check the status information, and fetch the results if they are ready. As both the result and the status information are small (e.g., one byte is enough to indicate whether the result is ready or not), they can be packed together for a client to fetch remotely. In this way, the client does not need to issue separate RDMA-read operations to fetch the status and the result. One RDMA-read is enough to fetch both items. This can reduce the number of round-trips between the client and the server and improve the IOPS.



\section{Design of An In-Memory Key-Value Store Following RFP}\label{design}

\begin{figure}[!t]
\centering
\includegraphics[width=0.9\columnwidth]{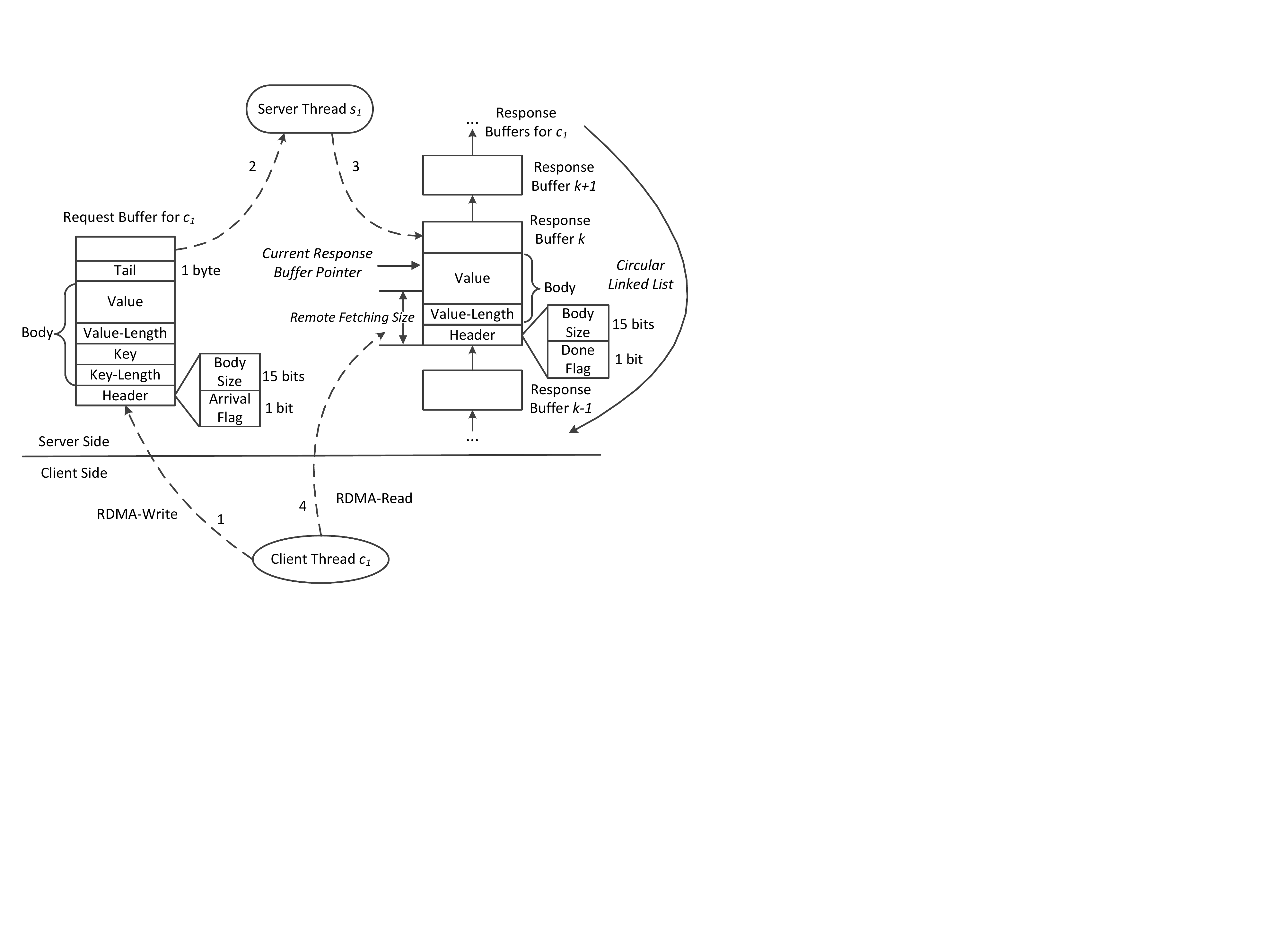}
\vspace{-0.5em}
\caption{The key data structures and procedures of the communication protocol in Jakiro.}
\vspace{-1.2em}
\label{j_protocol}
\end{figure}

Based on the Remote Fetching Paradigm, we design Jakiro, an RDMA-accelerated in-memory key-value store. Jakiro contains two important modules: (1) the underlying in-memory key-value structure and (2) the communication protocol to send requests (using RDMA-write) and remotely check/fetch results (using RDMA-read). We currently build the key-value store in cache mode, which can act as an important middleware to accelerate operations by keeping the items in memory like Memcached \cite{jakiro_memcached}. The key-value structure of Jakiro is partitioned across different server threads in Exclusive Read Exclusive Write (EREW) similar to MICA \cite{jakiro_mica}. As proved in previous work of \cite{jakiro_mica,jakiro_herd}, such design is able to provide high performance for processing key-value items. The communication protocol of Jakiro does also work well for other types of in-memory key-value structures such as Cuckoo hash \cite{jakiro_pilaf} or Hopscotch hash \cite{jakiro_farm}. We do not further discuss these key-value structures in this paper due to space limitation.

The communication protocol following RFP is designed as shown in Figure~\ref{j_protocol}. The protocol runs in both the server and the clients, and contains the following three procedures:

\textbf{Sending Requests.} The client first decides to which server thread it should send the request by the hash value of the key. Then it fills the request in the body of the local request buffer, and prepares the header of this buffer. At last, it appends a one-byte tail to the end of the body and sets it as 1. When these operations are done, the client thread sends the request buffer to the server thread by using RDMA-write.

\textbf{Processing Requests.} The server thread periodically checks the local request buffer for each client thread. It gets the request when it finds the \emph{arrival flag} in the header as well as the tail both become 1. Then it processes the request, clears the request buffer, and puts the result in the local response buffer. In Jakiro, each server thread uses a \emph{Current Response Buffer Pointer} (CRBP) to point to the response buffer that could be used for current request. Before setting the \emph{done flag} of the header, the server thread clears the two-byte header of the buffer \emph{next to current response buffer} in advance. This guarantees no old status or results will be fetched by the client thread for new requests. When the result is successfully put into current buffer, CRBP will be moved to the next buffer.

\textbf{Fetching Results Remotely.} After sending a request to the server, the client thread remotely checks the header of the response buffer maintained at the server side. The client thread also maintains a CRBP locally and selects the response buffer based on it. We set a \emph{Remote Fetching Size} (RFS) for the client thread to check and fetch the result remotely. In this case, the client thread directly fetches a data region of RFS from the response buffer of the server into the local memory using RDMA-read. The data region contains the two-byte header as well as a part of the body. If the client thread finds the done flag is set, it then gets the body size from the header. As long as the body size is no larger than $RFS-2$, the remote fetching action successfully ends. This is because the value is encapsulated in the data region when remote fetching is performed. Such an optimization mechanism combines status checking and result fetching. Thus, it can help make the most of in-bound RDMA-read IOPS, as separated RDMA-read operations to fetch status and result are avoided. However, if the body size is larger than $RFS-2$, another RDMA-read operation is required to get the remaining data.

\section{Evaluation}\label{evaluation}

We have implemented Jakiro in C++ (about 2500 lines of code). The libraries used for RDMA verbs are \emph{rdmacm} and \emph{ibverbs} provided by Mallanox OpenFabrics Enterprise Distribution \cite{jakiro_mallanox}. Our experimental goal is to answer the following questions:
\begin{enumerate}
\item Does Jakiro perform well in terms of IOPS and latency on small key-value data?
\item Does Jakiro outperform other in-memory key-value stores that used server-reply and totally-bypass design paradigms?
\item Is Jakiro suitable to diverse key-value workloads?
\end{enumerate}

\subsection{Experimental Setup}

We use a cluster based on InfiniBand for the evaluation. The cluster contains eight machines, each of which is equipped with dual 8-core CPUs (Intel Xeon E5-2640 v2, 2.0 GHz), 96 GB memory space, and a Mallanox ConnectX-3 InfiniBand NIC (MT27500, 40 Gbps). All of these machines are connected by an 18-port Mallanox InfiniScale-IV switch. The machines run Ubuntu 14.04, with the MLNX-OFED-LINUX-2.3-2.0.0 driver provided by Mellanox for Ubuntu 14.04 \cite{jakiro_mallanox}. We use one machine as the server and other seven machines as the clients to run Jakiro. Four threads are launched in the server machine, which are enough to serve clients' requests even when the server's INIC is saturated by the clients. For each client thread, the server thread maintains 8 response buffers. Each client thread also maintains the same number of response buffers in its local memory accordingly.

\begin{figure}[!t]
\begin{minipage}{0.45\linewidth}
  \centerline{\includegraphics[width=1.9in]{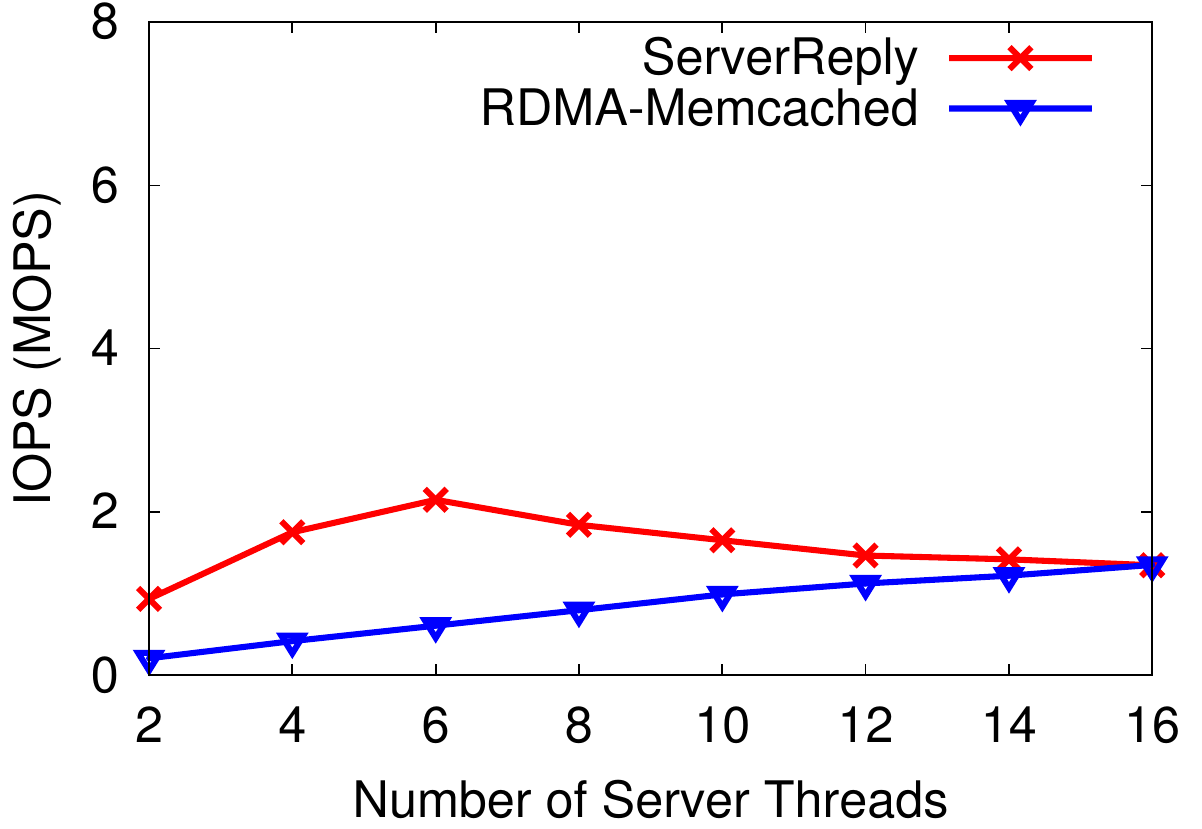}}
  \centerline{(a)}
\end{minipage}
\hfill
\begin{minipage}{0.45\linewidth}
  \centerline{\includegraphics[width=1.9in]{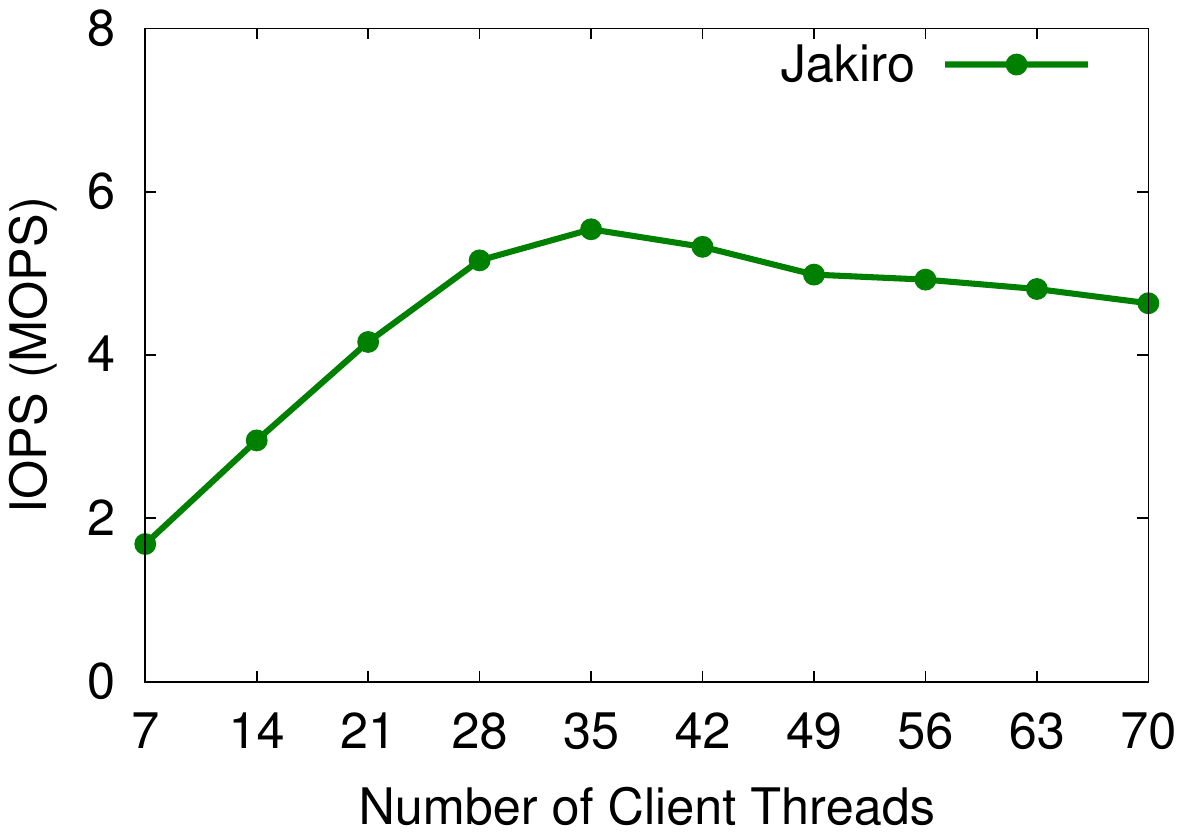}}
  \centerline{(b)}
\end{minipage}
\caption{The IOPS of Jakiro, ServerReply, and RDMA-Memcached on key-value items with value size of 32 bytes. The workload is uniform with 95\% GET (read-intensive).}
\vspace{-1em}
\label{j_iops}
\end{figure}

\textbf{Workloads.} We choose small key-value items with 16-byte key and 32-byte value. This is in line with the real-world workloads on small data
items according to the analysis of previous works \cite{jakiro_facebook,jakiro_herd,jakiro_smalldata}. Firstly, we present results on uniformly distributed and read-intensive workloads (95\% GET and 5\% PUT) in Section 5.2 and 5.3. Then we further provide the experimental results on skewed workloads and write-intensive workloads respectively.


We use YCSB to uniformly generate 128 million key-value items off-line for the experiment. Each item is accessed 20 times. The skewed workloads is generated according to Zipf distribution with parameter .99. For GET requests, we set RFS in Jakiro as 36 so as to encapsulate the 32-byte value in the remote fetching action \footnote{The remaining 4 bytes are used to store the two-byte header and the two-byte value-length, as shown in Figure~\ref{j_protocol}.}. We also use a real workload collected from Weibo service \cite{jakiro_weibo} to test Jakiro. This workload consists of 50,000,000 texts published by 93,633 users. The size of these texts ranges from 1 byte to 899 bytes and the average size is 43 bytes. The number of texts that have size larger than 512 bytes is only 331. We use the md5 value of user id and the time when publishing the text (i.e., \emph{md5(user\_id+time)}) as the key, and the original text as the value.


\textbf{Comparison.} We compare \Jakiro with two in-memory key-value systems. The first is \sr, which is extended from \Jakiro by using the traditional server-reply design paradigm. In other words, \sr differs from \Jakiro in that the server thread directly sends the result back to the client thread through RDMA-write after it completes the request. The other system is RDMA-based Memcached, which is developed by OSU \cite{jakiro_memcachedrdma} and also based on the server-reply design paradigm. We denote it as \osu. In \osu, the server thread will send status or notification information to the client thread after it processes the requests, and the client thread relies on these information to do further operations \cite{jakiro_memcachedrdma}. We run \osu in memory mode without interacting with the underlying permanent storage.

\subsection{Comparison on IOPS}
\begin{figure}
  \centerline{\includegraphics[width=2.85in]{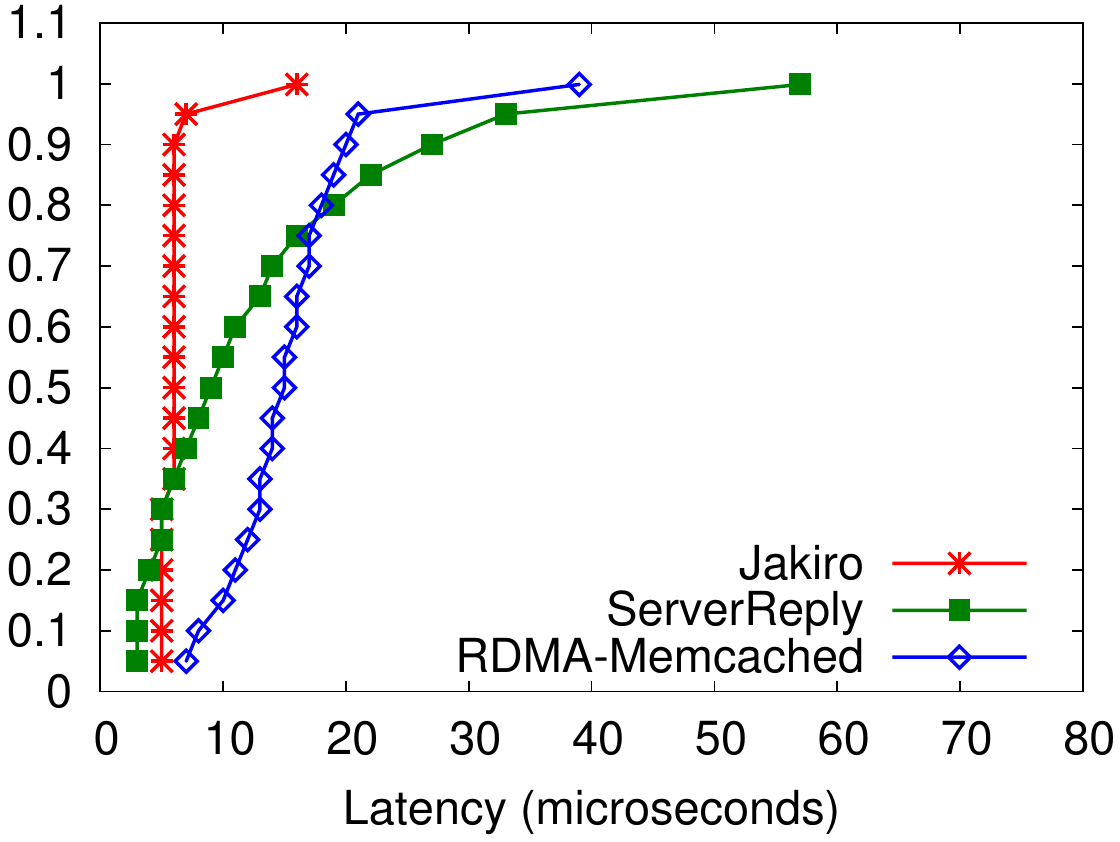}}\vspace{-0.5em}
  \caption{CDF of latency of Jakiro, ServerReply, and RDMA-Memcached on value size of 32 bytes.}\label{j_lat}
\end{figure}

\begin{figure}
  \centerline{\includegraphics[width=2.85in]{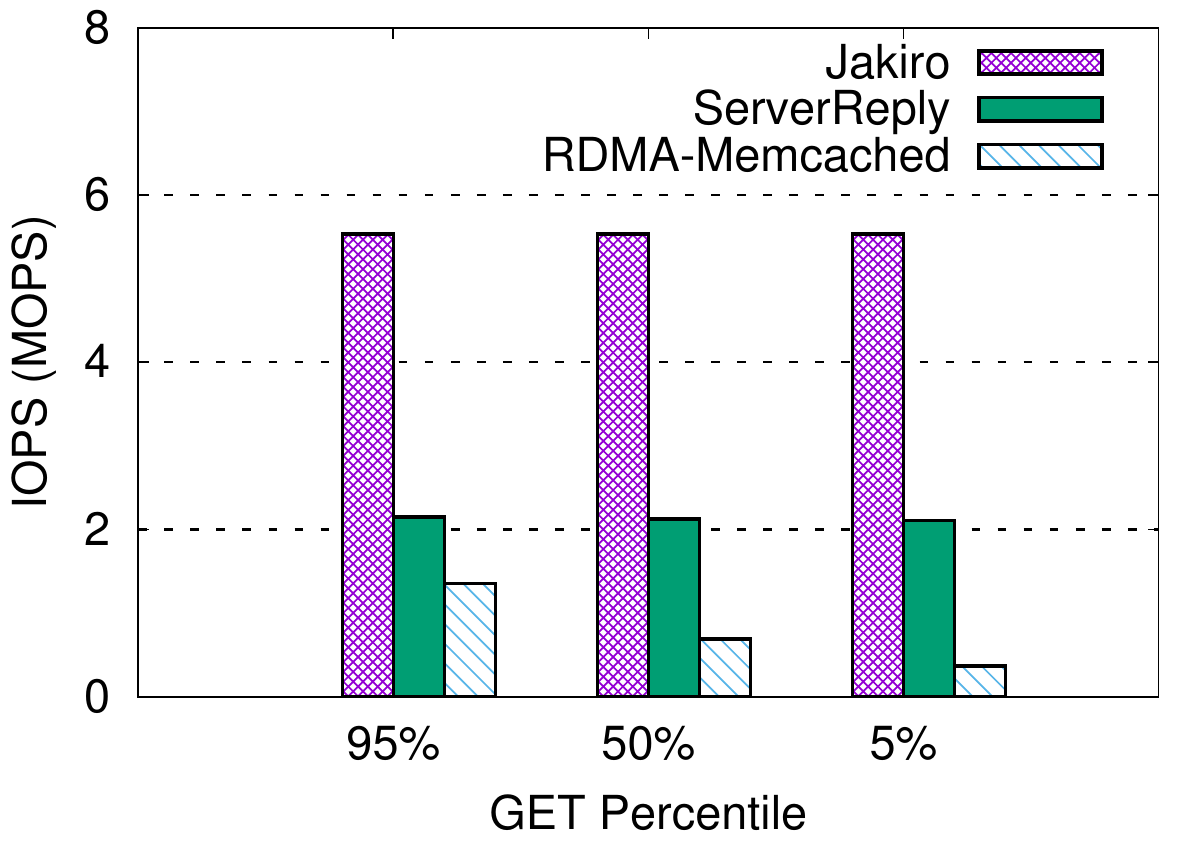}}\vspace{-0.5em}
  \caption{Comparison by varying GET percentiles (value size of 32 bytes under uniform workload).}\label{j_different_get}
\end{figure}

We first compare the three systems in terms of IOPS. As shown in Figure~\ref{j_iops}, the peak IOPS of \Jakiro on 32 bytes value is about 5.5 MOPS. Such peak IOPS is achieved when 35 client threads (uniformly distributed among seven client machines) are connecting to the server machine. We can observe that the peak IOPS is about half of the peak IOPS of pure in-bound RDMA-read for server's INIC (11.2 MOPS, as illustrated in Figure~\ref{j_iops_dif}). Thanks to the data encapsulation mechanism adopted in \Jakiro, two round-trips (2.005 round-trips on average for the evaluation) are needed for a client thread to successfully complete a request on small key-value items: one is to send the request through RDMA-write and the other is to fetch the result through RDMA-read. Thus, RFP can help \Jakiro make the most of INIC's capacity to achieve excellent IOPS.

The peak IOPS of \Jakiro is about 158\% higher than that of \sr (2.1 MOPS) and about 310\% higher than that of \osu (1.3 MOPS) respectively. The peak IOPS of \sr is achieved when 6 threads in the server are connecting to 28 client threads, while that of \osu is achieved at 16 server threads and 21 client threads. All the client threads are uniformly distributed among the client machines. Although \sr only needs two round-trips to complete a request (one is to send the request by the client thread and the other is to reply by the server), its peak IOPS on small key-value items is limited by the INIC's out-bound RDMA-write. Moreover, when the number of server threads increases, the IOPS of \sr decreases, due to the poor scalability of the INIC's out-bound RDMA operations.

It is interesting to see from Figure~\ref{j_iops}(a) that \osu is bounded by the CPU-utilization, and only increasing the number of server threads could improve its serving ability. However, even 16 server threads of \osu still cannot saturate the INIC's out-bound capacity, and thus brings lower peak IOPS than \sr. This is because a server thread of \osu has to coordinate with other threads for sharing data structures (e.g, LRU lists) as well as to perform network operations, which does not exhibit good scalability \cite{jakiro_pilaf,jakiro_memc3}. By using data partition, a server thread in \sr dose not need to interact with other threads, so \sr is only limited by out-bound RDMA operations. Moreover, as the server thread in \osu fully uses a CPU core, launching more than 16 server threads will cause the IOPS reduction for the reason that the server machine only has 16 physical CPU cores. Note that with the number of client thread further increased, the IOPS of \Jakiro decreases slightly, as shown in Figure~\ref{j_iops}(b). This is because when each client machine launches more client threads, the out-bound IOPS of its INIC will limit the number of operations it can perform in sending requests and fetching results remotely. However, despite of launching 112 client threads (each machine holds 16 threads), \Jakiro can achieve the IOPS of 3.1 MOPS. It is still higher than the peak IOPS of \sr and \osu respectively.





\subsection{Comparison on Latency}


The average latency on items with 32-byte value of \Jakiro is 5.78$\mu s$. It beats \sr's average latency (12.06$\mu s$) by 108\% and \osu's average latency (14.76$\mu s$) by 155\%. Figure~\ref{j_lat} illustrates the cumulative probability distribution of latency of the three systems when all of them achieve peak IOPS in the uniform and read-intensive workload. We can see from the figure that \sr has lower 15-percentile latency than \Jakiro. This is because a single RDMA-write has better performance than a single RDMA-read, as RDMA-write needs less state and operations than RDMA-read in the INIC hardware. Such phenomenon also has been observed in HERD \cite{jakiro_herd} and RDMA-PVFS \cite{jakiro_pvfs}. However, as the INIC has limitation on out-bound RDMA-write, \sr imposes higher latency than \Jakiro when more operations are observed (from 50-percentile to 100-percentile). In \Jakiro, about 99\% requests are below 7$\mu s$, which is significantly better than \sr (45\%) and \osu (10\%). Additionally, we can observe from Figure~\ref{j_lat} that all the three RDMA-bases systems suffer from the long-tail latency issue. To \Jakiro, some requests suffer higher latency (15$\sim$17$\mu s$) as they have to go through more round-trips (4\verb|-|6) for request sending and result fetching. However, the requests that have more than 2 round-trips only account for a small proportion (0.2\%) and will not impact the whole performance of \Jakiro.


\subsection{Evaluation with Different Workloads}

We compare the three systems under different types of workloads. We first report the IOPS of \Jakiro under different GET/PUT ratio. Then we vary the value size of key-value items for the testing. We give the performance of \Jakiro in skewed workload in the following. Finally, we report the results for a real workload. The experiment for the three systems is run at the same configurations when each of them can achieve peak IOPS on value size of 32 bytes in uniform and read-intensive (95\% GET) workload, as presented in Section 5.2. For \Jakiro, the configuration is 4 server threads connecting to 35 client threads. For \sr, the configuration is 6 server threads connecting to 28 client threads. For \osu, the configuration is 16 server threads connecting to 21 client threads. All the client threads are uniformly distributed among the machines.
\begin{figure}
  \centerline{\includegraphics[width=2.85in]{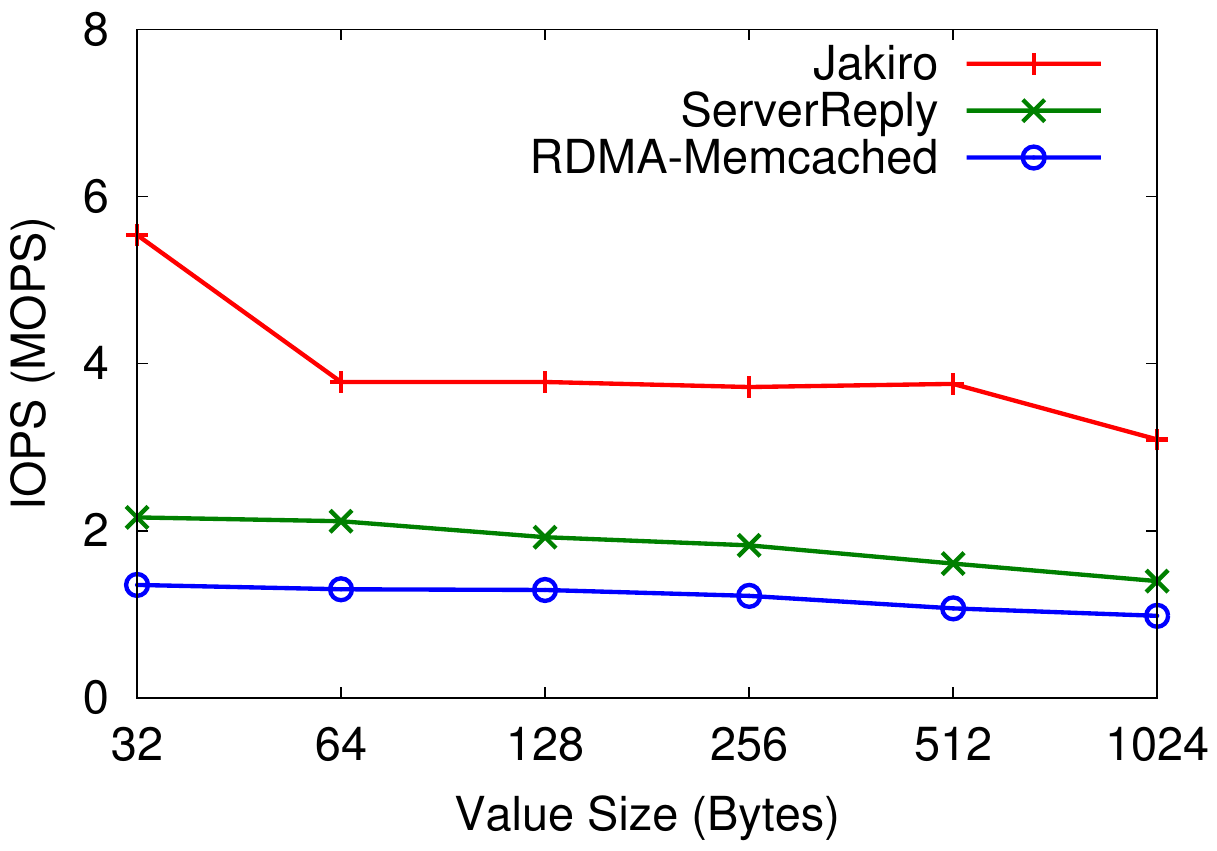}}\vspace{-0.5em}
  \caption{Comparison by varying value size. The workload is uniform with 95\% GET requests.}\label{j_value_size}
\end{figure}

\begin{figure}
  \centerline{\includegraphics[width=2.85in]{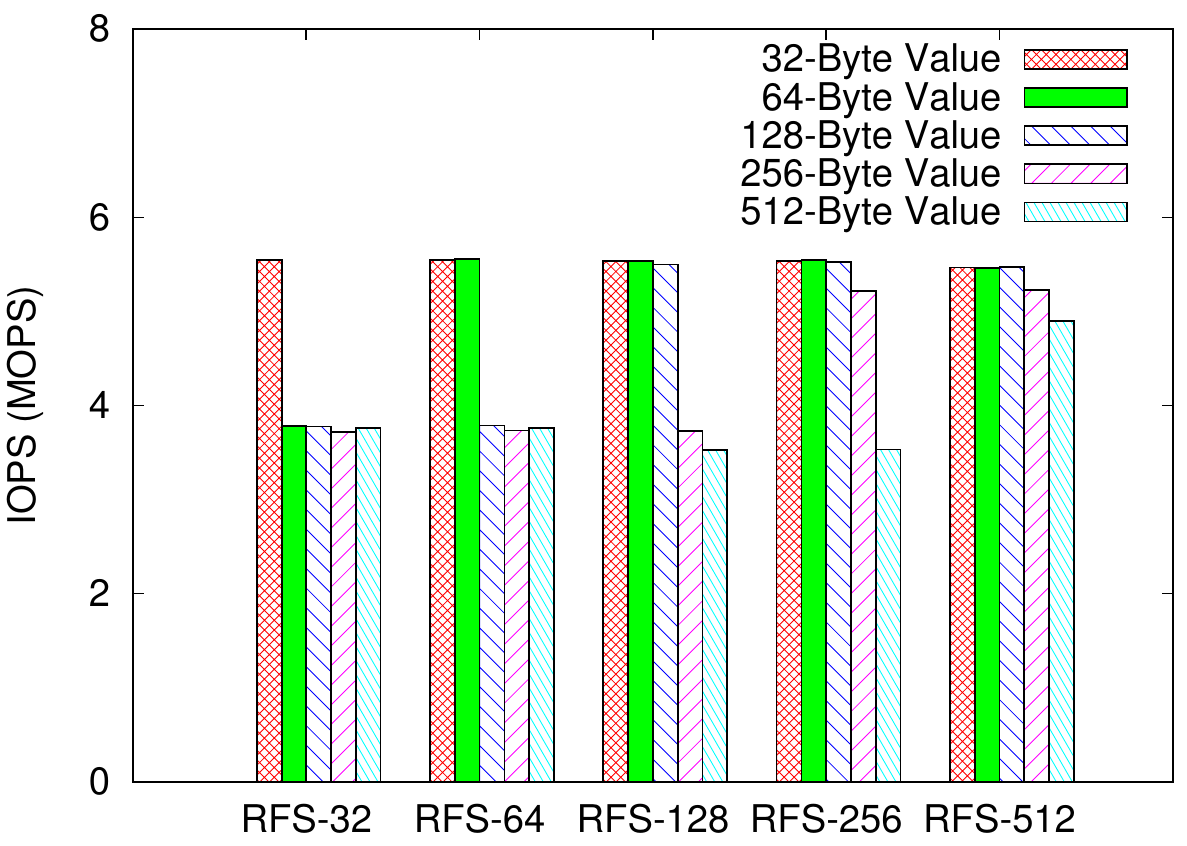}}\vspace{-0.5em}
  \caption{IOPS achieved by Jakiro under different Remote Fetching Size (RFS).}\label{j_rfs}
\end{figure}

\subsubsection{Varying GET Percentile}



Figure~\ref{j_different_get} illustrates the IOPS on 32-byte value size of \Jakiro, \sr, and \osu under different GET percentiles with uniform workload. \Jakiro can still fully use the INIC's performance to obtain peak IOPS at 5.5 MOPS under varying GET percentiles. As the server threads in \Jakiro are free from performing networking operations, their computing capacity is enough to process requests whether they are GET or PUT ones. That is why the peak IOPS of \Jakiro can reach 5.5 MOPS even when the workload is write-intensive (95\% PUT). Such IOPS is still about 160\% higher than that of \sr, which can saturate server INIC's out-bound RDMA-write performance (2.1 MOPS) under GET percentiles from 95\% to 5\%. However, as shown in Figure~\ref{j_different_get}, \osu is limited by the CPU utilization \footnote{The CPU utilization consists of computing, accessing memory, and performing network operations.}, and has decreasing IOPS when the requests are becoming write-intensive. In workload with 95\% PUT requests, \Jakiro improves the IOPS by 14X compared to \osu.


\subsubsection{Varying Value Size}

Figure~\ref{j_value_size} presents the IOPS of \Jakiro, \sr, and \osu on varying value size, under uniform workload with 95\% GET. It can be seen from Figure~\ref{j_value_size} that \Jakiro outperforms \sr and \osu on value size from 32 bytes to 1024 bytes. When the value size becomes larger, the server thread in \sr has to spend more time sending a result back to the client thread through out-bound RDMA-write for a GET request. Therefore, such IOPS decreasing does occur for \sr due to wasting cycles on networking operations. Increasing the number of server threads could help mitigate the issue for \sr on larger value size. For example, when the server thread number is increased to 12, \sr can achieve about 2.1 MOPS for key-value items with value size of 512 bytes. However, as shown in Figure~\ref{j_iops_dif}, it is at the expense of reducing IOPS on smaller value size (e.g., 32 bytes or 64 bytes). Therefore, traditional server-reply design paradigm does not work well for relatively small requests ($\sim$1024 bytes) over RDMA. On the contrast, the server thread in \Jakiro does not have to spend cycles in networking communication, and thus can process more requests. This further proves the effectiveness of RFP in optimizing IOPS on small data by leveraging the INIC's excellent in-bound RDMA-read performance.


In Figure~\ref{j_value_size}, we can observe that when the value size is from 64 bytes to 512 bytes, the IOPS is about 32\% lower than that on value size of 32 bytes in \Jakiro. This is because the RFS of \Jakiro is set 36 (as mentioned in Section 5.1) for GET requests, and the value with size larger than 32 bytes cannot be totally encapsulated in the data region remotely fetched by the client thread. When the client thread checks the result status is ready, it has to issue another RDMA-read operation to fetch the remaining data if it finds the body size in the response buffer's header is larger than $RFS-2$. As a result, three round-trips on average are needed if the key-value item has too large value size to be encapsulated in the data region of remote fetching action. The 32\% reduction of IOPS on value size from 64 bytes to 512 bytes is reasonable when RFS is set 36, as two round-trips has about 33\% reduction compared to three round-trips. Increasing RFS to encapsulate the whole data in the remote fetching actions can help improve the IOPS of larger value size. For example, as illustrated in Figure~\ref{j_rfs}, when setting RFS as 132 (128 bytes for value plus 4 additional bytes \footnote{The X-axis in Figure~\ref{j_rfs} only reflects the value size and does not include the 4 additional bytes.}), the IOPS on key-value items with value size of 32 bytes, 64 bytes, and 128 bytes can all reach around 5.5 MOPS due to round-trip reduction. Note that when the value size grows to 1024 bytes, the IOPS of \Jakiro is smaller (around 3.1 MOPS, according to Figure~\ref{j_value_size}) as the network's bandwidth is saturated and becomes the bottleneck in this case.




\subsubsection{Skewed Workload}

\begin{figure}
  \centerline{\includegraphics[width=2.85in]{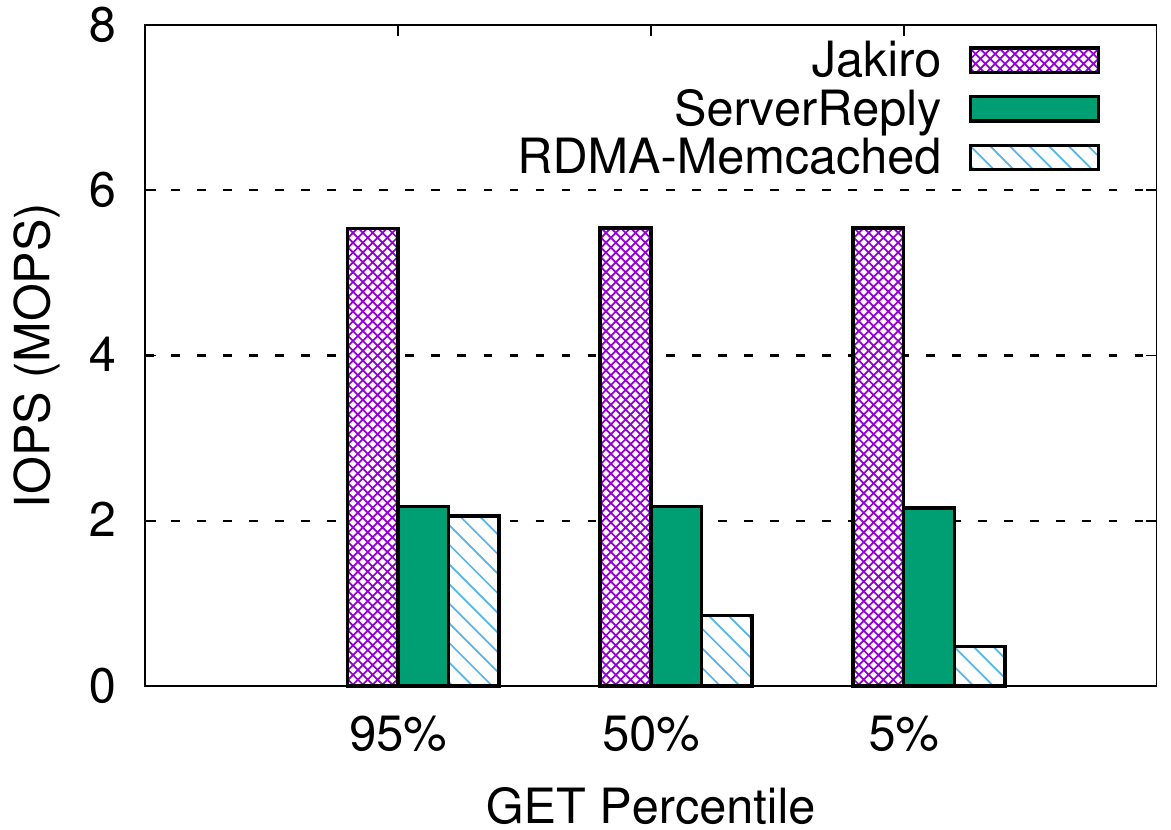}}\vspace{-0.5em}
  \caption{Comparison with skewed workloads according to Zipf with parameter .99 (32-byte value size).}\label{j_skew}
\end{figure}

We test how \Jakiro perform in a skewed workload. The keys in the workload are generated according to a Zipf distribution with parameter .99. Figure~\ref{j_skew} presents the IOPS of \Jakiro, \sr, and \osu on value size of 32 bytes under different GET percentiles. Although the most popular key is about $10^{5}$ times more than the average key in the skewed workload, the most loaded server thread is $<$20\% more than the thread with the least load \cite{jakiro_herd}, in the case of launching four server threads. Even under skewed workload, the server threads in \Jakiro are able to process small key-value requests when the server's INIC is saturated. Thus, the peak IOPS of \Jakiro is still 5.5 MOPS under 5\%, 50\%, and 95\% GET percentiles. Similarly, \sr is still limited by the INIC's out-bound RDMA-write instead of the CPUs, and obtains 2.1 MOPS for the skewed workload. As mentioned in Section 5.2 and Section 5.4.1, the \osu is bounded by the CPU at the server side. Under skewed workload, \osu could benefit from serving the popular keys as this will make use of cache locality \cite{jakiro_chint}. As a result, the IOPS of \osu in this case is higher than that under uniform workload. For example, as shown in Figure~\ref{j_skew}, \osu achieves about 2.1 MOPS on value size of 32 bytes with 95\% GET requests, which saturates the INIC's capacity for out-bound RDMA. Overall, \Jakiro also beats both \sr and \osu under workload with skewed distribution.

\subsubsection{Real Workload}



\begin{figure}
  \centerline{\includegraphics[width=2.85in]{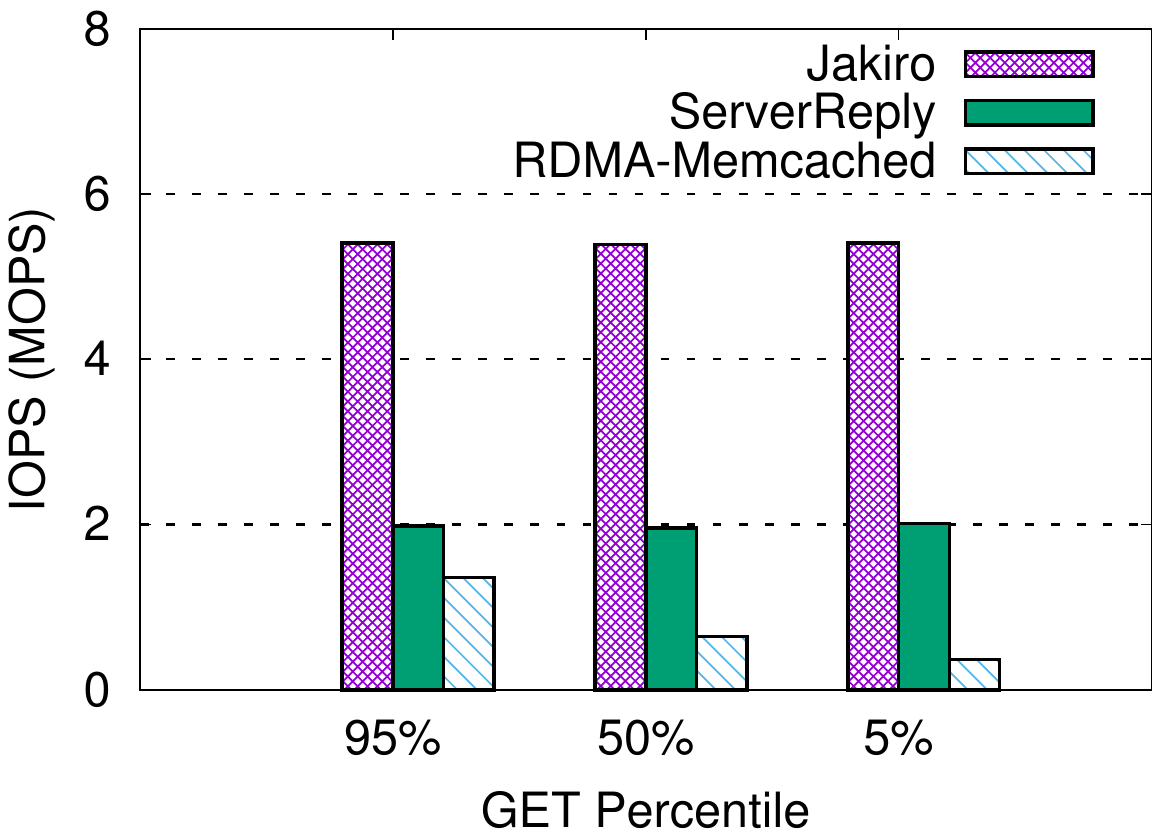}}\vspace{-0.5em}
  \caption{Comparison on a real workload from Weibo service under different GET percentiles.}\label{j_weibo}
\end{figure}

We use a real workload collected from a popular Weibo service \cite{jakiro_weibo} to test \Jakiro. The RFS of \Jakiro is set 512. Figure~\ref{j_weibo} plots the IOPS gotten by \Jakiro, \sr, and \osu for the real workload. We can see from this figure that the peak IOPS of \Jakiro under different GET percentiles is 5.4 MOPS, which is still much higher than that of \sr and \osu. This further verifies the effectiveness of RFP in optimizing the IOPS for systems.

\subsection{Discussion}

As mentioned in Section 1, there is another design paradigm to apply RDMA to applications, i.e., totally bypassing server CPUs to complete requests. \Pilaf \cite{jakiro_pilaf} is a state-of-the-art in-memory key-value store by using such totally-bypass design paradigm. However, as clients gets no information about whether the corresponding item is modified or moved at the server side, more round-trips are needed for completing a request. This will significantly impact the IOPS on small data. The peak IOPS of \Pilaf on 64-byte values reported in the work of \cite{jakiro_pilaf} is only 1.3 MOPS (90\% GET) over 20 Gbps Mallanox INICs. We also run \Jakiro in a cluster of six machines equipped with 20 Gbps Mallanox INICs, the same as the INICs tested for Pilaf. The peak IOPS of \Jakiro on 64-byte values with 90\% GET is about 5.4 MOPS, which is about 4x as high as that of \Pilaf.

Additionally, it is complex to apply the totally-bypass design paradigm in more sophisticated mechanisms such as batching or transaction management, which need to operate a number of data items located in different memory addresses of the server. Although Paxos \cite{jakiro_paxos} can be used among the clients to complete batch or transaction requests, this further complicates the design and implementation and imposes even worse performance. In contrast, RFP can effectively handle these requests by making server CPUs process the requests and clients remotely fetch the results. Server CPUs have the performance and locality advantage of processing such requests.


\section{Related Work}\label{related_work}
We review related works, and show the similarity and difference between our work and them in this section.

\textbf{Totally-Bypass Design Paradigm.} Most studies have attempted to leverage the features of RDMA verbs to totally bypass server CPUs in the client-server architecture to achieve high performance. For example, Pilaf \cite{jakiro_pilaf} allows its clients to directly read data from the server's memory through RDMA-read for GET requests. The clients use CRC64 to check for data inconsistency caused by potential races with the server. C-Hint \cite{jakiro_chint} designs an in-memory key-value store by using RDMA similar to Pilaf. It focuses on cache management issues such as tracking popular data, making replacement decisions, and reclaiming memory resources safely in this case. FaRM \cite{jakiro_farm} is a memory distributed computing platform that utilizes RDMA to achieve higher performance than TCP/IP. It also uses RDMA-read to perform its lock-free reads without involving the server.


However, the totally-bypass design paradigm has disadvantages in supporting general-purpose systems such as key-value stores or RPC servers. As the server is bypassed, many policies (e.g., access pattern counting, data race avoiding, or data validity checking) that could have been executed locally and efficiently in the server must be ported to the remote clients in a more complex way. In most cases, this needs tremendous re-design of upper-level systems and is inefficient. For example, Pilaf, C-Hint, and FaRM have to propose solutions to reason about data consistency. C-Hint has to waste extra networking resources for the interaction between clients and the server to count access patterns. Moreover, as the clients have no information about how data are moved or replaced in the server, many round-trips may be needed to locate and fetch a data item. This will cause the reduction of performance. Compared to this design paradigm, Remote Fetching Paradigm proposed in this paper has the general feature by making the server involved in the process and can make the most of the INIC's performance. For instance, the IOPS of Jakiro on small data is about 4X as high as that of Pilaf \cite{jakiro_pilaf}. It has to be mentioned that to Pilaf, C-Hint, and FaRM, all of them use server-reply design paradigm to serve PUT requests. In this case, these systems will also suffer from the limited IOPS of server's out-bound RDMA on small data.

\textbf{Server-Reply Design Paradigm.} There are also many previous works have applied RDMA to systems such as Memcached \cite{jakiro_memcachedrdma}, HDFS \cite{jakiro_hdfs}, PVFS \cite{jakiro_pvfs}, YARN \cite{jakiro_yarn}, HBase \cite{jakiro_hbase}, etc., by using traditional server-reply design paradigm. Although it can generally support any client-server system, we argue that in such design paradigm, RDMA cannot improve IOPS on small data much due to the limited out-bound RDMA performance in the server. Compared to server-reply design paradigm, Remote Fetching Paradigm can better exploit more scalable and attractive in-bound RDMA performance by making clients fetch results remotely, and is able to provide higher IOPS (as proved in Section~\ref{evaluation}).

HERD \cite{jakiro_herd} is a state-of-the-art in-memory key-value store that efficiently uses server-reply design paradigm. However, it relies on unreliable transport services, i.e., Unreliable Connection (UC) and Unreliable Datagram (UD), to achieve high IOPS. Systems that require reliability \cite{jakiro_chint,jakiro_pilaf,jakiro_farm,jakiro_pvfs,jakiro_memcachedrdma,jakiro_wimpy} cannot benefit from such transport services without reliability guarantee. Realizing reliability upon the unreliable services at the software level not only complicates the system or application design \cite{jakiro_chint}, but cannot outperform purely using Reliable Connection (RC), a reliable transport service over InfiniBand \cite{jakiro_memcachedrdma,jakiro_scalablememcachedrdma}. Moreover, RC can support all RDMA verbs while UC does not support RDMA-read and UD does not support any RDMA-relevant operation. In RFP, all RDMA operations are based on RC.



\section{Conclusions}\label{conclusion}
This paper proposes a novel design paradigm named Remote Fetching Paradigm (RFP) to improve the RDMA-accelerated client-server systems. By making server CPUs involving in request processing, RFP is able to generally support various systems or applications. Moreover, RFP can achieve much higher IOPS on small data by counter-intuitively using clients to check and fetch results actively and remotely. This can make the most of server's in-bound RDMA performance. We have designed and implemented an in-memory key-value store named Jakiro following the model of RFP. The experimental results show that Jakiro outperforms in-memory key-value stores following totally-bypass and server-reply design paradigms on small data. We believe RFP can help researchers and developers to rethink the design of traditional RDMA-accelerated client-server systems.




\end{document}